# Infant movement classification through pressure distribution analysis – added value for research and clinical implementation

*Tomas Kulvicius,[1,2][*][~], Dajie Zhang[1,3,4][*], Karin Nielsen-Saines[5], Sven Bölte[6,7,8], Marc Kraft[9], Christa Einspieler[4], Luise Poustka[1,3], Florentin Wörgötter[3,9+], Peter B Marschik[1,3,4,6+]*

1. Child and Adolescent Psychiatry and Psychotherapy, University Medical Center Göttingen, Göttingen, Germany
2. Department for Computational Neuroscience, Third Institute of Physics-Biophysics, Georg-August-University of Göttingen, Göttingen, Germany
3. Leibniz-ScienceCampus Primate Cognition, Göttingen, Germany
4. iDN – interdisciplinary Developmental Neuroscience, Division of Phoniatrics, Medical University of Graz, Graz, Austria
5. Division of Pediatric Infectious Diseases, David Geffen UCLA School of Medicine, Los Angeles, USA
6. Center of Neurodevelopmental Disorders (KIND), Centre for Psychiatry Research; Department of Women's and Children's Health, Karolinska Institutet, Stockholm, Sweden
7. Child and Adolescent Psychiatry, Stockholm Health Care Services, Region Stockholm, Stockholm, Sweden
8. Curtin Autism Research Group, Curtin School of Allied Health, Curtin University, Perth, Western Australia
9. Department of Medical Engineering, Technical University Berlin, Berlin, Germany
   ~Corresponding author (E-mail: tomas.kulvicius@uni-goettingen.de)
   *These authors share first authorship
   [+]These authors share senior authorship

**Abstract**

Aiming at objective early detection of neuromotor disorders such as cerebral palsy, we proposed an innovative non-intrusive approach using a pressure sensing device to classify infant general movements (GMs). Here, we tested the feasibility of using pressure data to differentiate typical GM patterns of the "fidgety period" (i.e., fidgety movements) vs. the "pre-fidgety period" (i.e., writhing movements). Participants (N = 45) were sampled from a typically-developing infant cohort. Multi-modal sensor data, including pressure data from a 32x32-grid pressure sensing mat with 1024 sensors, were prospectively recorded for each infant in seven succeeding laboratory sessions in biweekly intervals from 4-16 weeks of post-term age. For proof-of-concept, 1776 pressure data snippets, each 5s long, from the two targeted age periods were taken for movement classification. Each snippet was pre-annotated based on corresponding synchronised video data by human assessors as either fidgety present (FM+) or absent (FM-). Multiple neural network architectures were tested to distinguish the FM+ vs. FM- classes, including support vector machines (SVM), feed-forward networks (FFNs), convolutional neural networks (CNNs), and long short-term memory (LSTM) networks. The CNN achieved the highest average classification accuracy (81.4%) for classes FM+ vs. FM-. Comparing the pros and cons of other methods aiming at automated GMA to the pressure sensing approach, we concluded that the pressure sensing approach has great potential for efficient large-scale motion data acquisition and sharing. This will in return enable improvement of the approach that may prove scalable for daily clinical application for evaluating infant neuromotor functions.

## Introduction

Over the past decades, our knowledge on human spontaneous movements, their onset, developmental trajectory, and predictive value for clinical outcomes has become increasingly profound[1,2]. Among the rich presentations of infant movements, a specific spontaneous motor repertoire, termed general movements by Prechtl and colleagues[2,3], has gained substantial attention. The Prechtl general movements assessment (GMA) has proven to be an efficient and reliable diagnostic tool for detecting cerebral palsy within the first few months of human life[4]. The significance of general movements as a biomarker for divergent early brain development, and their long-term relevance for cognitive, speech-language, and motor development have been widely acknowledged[5-8].

GMA is a method based on visual gestalt perception. The expertise of the GMA assessors comes from intensive training and sustaining practice. The cumulative cost and effort required for human assessors to achieve and maintain adequate performance are considerable. In common with other man-powered assessments, human factors may affect GMA assessors' performance, which may have impeded an even broader application of this efficient diagnostic tool. As a consequence, increasing efforts on automated solutions to classify infant motor functions have been made in the last years to supplement the classic GMA[9-12], where most attempts have been, remaining true to the GMA methodology, devoted to developing vision-based solutions, for example see[13-18]. Several sensor-based methods directly capturing 3D motion data of the infants have also been developed[19-25]. Both vision-based methods using marker-based body tracking[26] and methods applying non-vision sensors[19-23], require attaching elements or devices to the infant's body, which can become cumbersome and might alter infants' behavioural status and their motor output[27]. To optimise tracking and evaluating developmental behaviours in early infancy, we developed in 2015 a comprehensive multimodal approach[28]. One of our aims was to systematically examine the potential of different techniques and their combinations to classify, among others, infant movement patterns. With a prospective longitudinal approach, we recruited a cohort of typically developing infants. For recording their movements, we utilised, in addition to a multi-camera video set-up (2D-RGB and 3D Kinects), inertial motion units (IMUs, accelerometer sensors), and a pressure sensing mat[28].

Pressure sensitive mats have been broadly applied in infant monitoring, sport training, and patient care to evaluate dynamic force distribution and displacement in sitting, lying, or standing positions in individuals with different mobilities and at different ages[29-39]. Pressure sensing devices have also been used for assessing preterm- and term-infants' sleeping behaviours, gross motor patterns, and postural control[40-50]. Reported in a recent abstract, Johnson and colleagues[51] used a force plate to assess motor patterns in 12 typically developing infants aged 2 to 7 months. They clustered the infants into three groups according to their movement variabilities (i.e., moderate, mild, and little variability). Greater variability captured by the pressure sensing device seemed to be associated with an age-specific general movement pattern, the fidgety movements (FMs), i.e., a general movement pattern that presents during 9-20 weeks of post-term age in typically developing infants[27]. No further technical details were revealed in the abstract. Kniaziew-Gomoluch and colleagues examined the postures of infants who presented either normal or abnormal fidgety movements. They found statistical differences between the two groups in their Centre of Pressure (CoP) parameters measured by a force plate[45,46]. The authors, however, did not apply machine learning to classify general movements.

As pressure sensing mats record force changes in motion across spatial and temporal dimensions, it has the potential to distinguish infant movements that are different in their timing, speed, amplitude, spatial distribution, connection, and organisation (reflecting the involvement and displacement of different body-parts), which are the characterising features discriminating general movements at different ages, and more importantly, of distinctive qualities[27] (physiological vs. pathological). Compared to other motion tracking techniques, the application of pressure mats does not require complex and time-consuming setups and is fully non-intrusive to the infants. The pressure mat data can be readily deidentified thus circumvent potential data privacy issues when it comes to data transfer and sharing[52]. If a pressure sensitive mat can reliably detect different infant general movement patterns, it will have great potential for routine clinical applications which may substantially increase the accessibility of GMA.

With the current proof-of-concept study, we sought to test the viability of using a pressure sensitive mat to classify different general movement patterns. We intended to use the pressure sensing mat to first analyse typical development, which shall build the basis for future investigations targeting altered development (i.e., classification of typical vs. atypical patterns pinpointing neurological dysfunction). Utilising data obtained from the aforementioned prospective-longitudinal infant cohort[28], we aimed to examine whether pressure data can be used to differentiate typical fidgety movements from "pre-fidgety movements", i.e., writhing movements[27].

**Methods**

For the current study, we used a validated expert-annotated dataset reported in a previous study of our research group[16] (please find details below; see also Figure 1). Data acquisition was done at iDN's BRAIN*tegrity* lab at the Medical University of Graz, Austria. Movement data were collected as a part of the umbrella project with a prospective longitudinal design aimed to profile typical cross-domain development during the first months of human life[28]. Data analyses for the current study were done at the Systemic Ethology and Developmental Science Unit - SEE, Department of Child and Adolescent Psychiatry and Psychotherapy at the University Medical Center Göttingen, Germany. The study was approved by the Institutional Review Board of the Medical University of Graz, Austria (27-476ex14/15) and the University Medical Center Göttingen, Germany (20/9/19). Parents were informed of all experimental procedures and study purpose, and provided their written informed consent for participation and publication of results.

**Participants**
Participants of the umbrella project[28] included 51 infants born between 2015 and 2017 to monolingual German-speaking families in Graz (Austria) and its close surroundings. Inclusion criteria were: uneventful pregnancy, uneventful delivery at term age (>37 weeks of gestation), singleton birth, appropriate birth weight, uneventful neonatal period, inconspicuous hearing and visual development. All parents completed high-school or higher level of education. The parents had no record of alcohol or substance abuse. From the 51 infants, one was excluded due to a diagnosed medical condition at 3 years of age. Another five were excluded due to incompleteness of recordings within the required age intervals (see below). The final sample for the current study comprised 45 infants (23 females).

**Data acquisition**
From 4 to 16 weeks of post-term age, each infant was assessed at 7 succeeding sessions in a standard laboratory setting in biweekly intervals. Post-term ages at the seven sessions were: T1 28 ± 2 days, T2 42 ± 2 days, T3 56 ± 2 days, T4 70 ± 2 days, T5 84 ± 2 days, T6 98 ± 2 days, and T7 112 ± 2 days. According to the GMA manual[27], 5 to 8 weeks of post-term period (T2 and T3 belong to this period) is a transitional- or "grey"-zone between the writhing and the fidgety periods and is not ideal for assessing infant general movements. Fidgety movements are most pronounced in typically developing infants from 12 weeks of post-term age onwards[27] (corresponding to T5-7). Therefore, to analyse infant general movements, data from T1 as "pre-fidgety period" and T5-7 as "fidgety period" were taken for the current study.

Infant movement data were recorded in form of RGB and RGB-D video streams, accelerometer and gyroscope data, and pressure sensing mat data[28]. All sensors were synchronised. Data recording procedure followed the GMA guidelines[27]. The pressure data was acquired using a Conformat pressure sensing mat[53] (Tekscan, Inc., South Boston, Massachusetts, USA). The mat was laid on the mattress, covered by a standard cotton sheet. During a laboratory assessment, the infant was placed in supine on the mat by the parent. The Conformat contains 1024 pressure sensors arranged in a 32 x 32 grid array on an area of 471.4 x 471.4 mm$^2$, producing pressure image frames (8 Bit, 32 x 32 pixels, sampling rate 100 Hz).

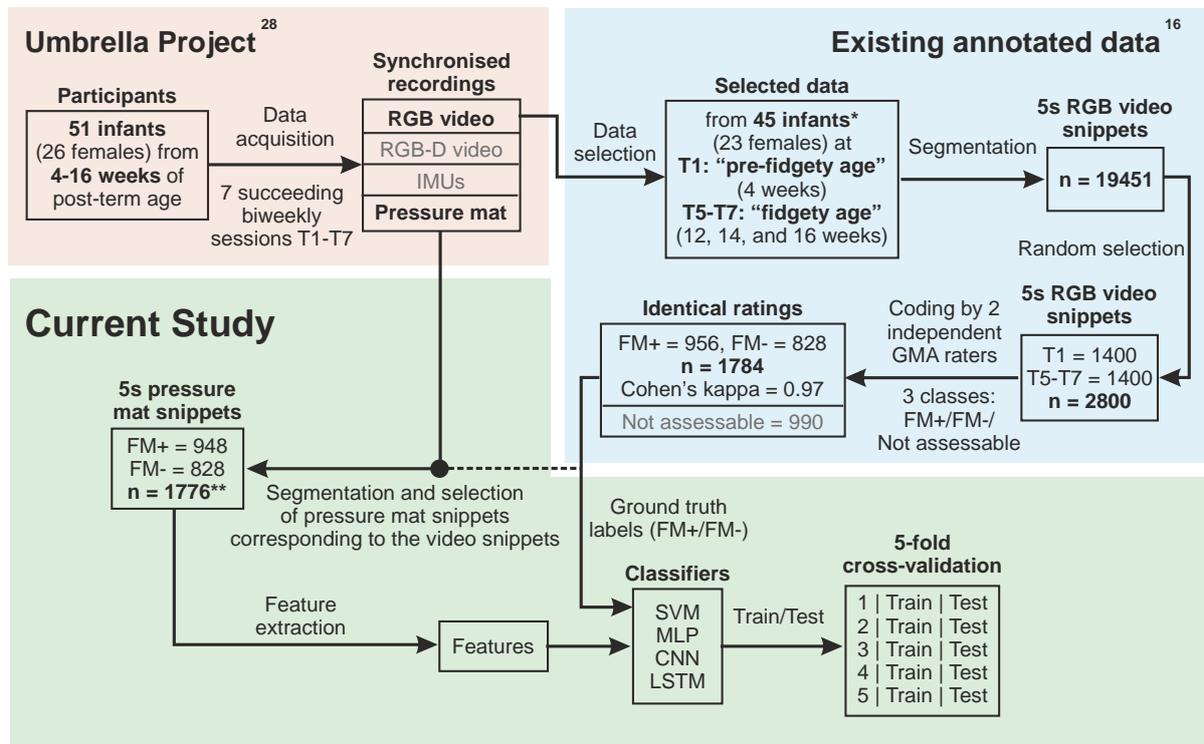

**Figure 1:** Diagram of the study pipeline. The numbers "n" correspond to the number of snippets in each step.

**Data annotation**

For movement classification using machine learning methods with pressure mat data, human-annotation data was needed. These annotation data were available from a previous study[16]. Human annotation was based on RGB recordings which were synchronised with the pressure mat recordings. In that previous study[16], we first cut suitable videos (i.e., infants were, overall, awake and active and not fussy) from T1 and T5-7 of the 45 infants each into brief chunks (i.e., snippets). Based on initial pilot trials, we determined the shortest length of each video snippet to be 5 seconds, a reasonable duration of unit for machine learning, as well as a minimum length of video for human assessors feeling confident to judge whether the fidgety movement is present (FM+) or absent (FM-) on each snippet[16], providing the ground truth labels to train and test classification models as described below.

For the purpose of proof-of-concept, only a fraction of the total available snippets (N = 19451) was sampled and annotated by human assessors. Out of the entire pool, 2800 snippets were randomly chosen: 1400 from T1, the pre-fidgety period, and the other 1400 from T5-7, the fidgety period. Two experienced GMA assessors, blind of the ages of the infants, evaluated each of the randomly ordered 2800 five-second snippets independently, labelling each snippet as "FM+", "FM-", or "not assessable" (i.e., the infant during the specific 5 seconds was: fussy/crying, drowsy, hiccupping, yawning, refluxing, over-excited, self-soothing, or distracted, all of which distort infants' movement pattern and shall not be assessed for GMA[27]. The interrater agreement of the two assessors for classes FM+ and FM- was Cohen's kappa κ = 0.97. The intra-assessor reliability by rerating 280 randomly-chosen snippets (i.e. 10% of the sample) was Cohen's kappa κ = 0.85 for assessor 1, and κ = 0.95 for assessor 2 for the classes FM+ and FM-. Snippets with discrepant labelling for classes FM+ vs. FM- by the assessors (N = 24), and the ones labelled as "not assessable" by either assessor (N = 990) were excluded. A remaining total of 1784 video snippets were labelled identically by both assessors as either FM+ (N = 956), or FM- (N = 828). Of the 1784 snippets, 1776 had corresponding synchronised pressure mat data, which were used for the machine learning procedures described below. Among the 1776 pressure mat snippets, 948 adopted the corresponding label of "FM+", and 828 of "FM-".

**Feature Extraction for Motion Encoding**

A flow diagram of the feature extraction procedure is shown in Figure 2. As input, we used the 1776 pressure mat recordings, each 5s long, corresponding to the 5s video snippets described above[16], with a sampling rate of 100 Hz, which led to 500 frames per snippet. One frame consists of 1024 pressure sensor values arranged on a 32x32 grid (see frames on the left side in Figure 2).

We first cropped the area [1:29, 4:29] of original grid size 32x32 (red rectangle), since in most cases the sensor values outside this area were 0. Thus, the size of the cropped area was 29x26 leading to 754 pressure sensor values. Generally, only two areas were strongly activated on the pressure mat, where activations on the top correspond to the infant shoulders and/or head, and activations at the bottom correspond to the infant hips. Therefore, we split the cropped grid of size 29x26 into two parts, 12x26 (top) and 17x26 (bottom), and tracked the centre of pressure (CoP) in these two areas.

Next, we computed position coordinates *x* and *y* of the CoP and the average pressure values *p* of the top and the bottom areas for each frame as the following:

$$x_{t/b} = \frac{\sum_i i \times p_{t/b}(i,j)}{\sum_{i,j} p_{t/b}(i,j)}, \quad y_{t/b} = \frac{\sum_j j \times p_{t/b}(i,j)}{\sum_{i,j} p_{t/b}(i,j)}, \quad p_{t/b} = \frac{\sum_{i,j} p_{t/b}(i,j)}{m_{t/b} \times n_{t/b}}.$$

Here, $p_t(i,j)$ and $p_b(i,j)$ correspond to the pressure sensor values at the position ($i=1..m_{t/b}$, $j=1..n_{t/b}$) of the top (t) and bottom (b) parts, respectively. To reduce signal noise, for each value *x*, *y*, and *p*, we applied moving average filter with a sliding window of size 5 frames (0.05 s).

To avoid biases that could be caused by infant size and weight, we normalised values *x*, *y*, and *p* between 0 and 1 as follows:

$$x_{t/b} = \frac{x_{t/b} - min(x_{t/b})}{max[max(x_t) - min(x_t), max(x_b) - min(x_b), max(y_t) - min(y_t), max(y_b) - min(y_b)]},$$
$$y_{t/b} = \frac{y_{t/b} - min(y_{t/b})}{max[max(x_t) - min(x_t), max(x_b) - min(x_b), max(y_t) - min(y_t), max(y_b) - min(y_b)]},$$
$$p_{t/b} = \frac{p_{t/b} - min(p_{t/b})}{max[max(p_t) - min(p_t), max(p_b) - min(p_b)]}.$$

Thus, the original input of size 500 x 32x32 was reduced to 500 x 6, i.e., six signals of 500 time steps.

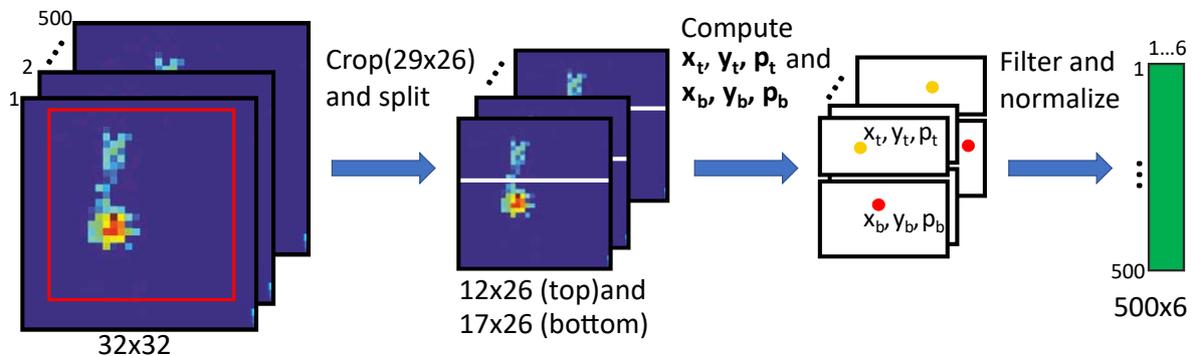

**Figure 2:** Flow diagram of the feature extraction procedure for the motion encoding.

**Classification models**

For classifying infant movements (FM+ vs. FM-), we compared a support vector machine (SVM) and a feed-forward network FFN, also known as multi-layer perceptron (MLP), with manually defined features against a convolutional neural network (CNN) and a long-short term memory (LSTM) network with learned features (for examples of network architectures see Figure 3).

In case of the SVM and FFN, we used statistical features obtained from signals $x_{t/b}$, $y_{t/b}$, and $p_{t/b}$ and their derivatives $x'_{t/b}$, $y'_{t/b}$, and $p'_{t/b}$ by computing mean and standard deviation values for each signal. In one case, we only used statistical features from the original signals, which resulted in 12 input values in total. In another case, we used statistical features from both the original signals and their derivatives, which resulted in 24 input values in total. As shown in Table S1, we investigated SVMs with different kernels (RBF and polynomial) and FFN architectures with one or two fully connected (FC) layers with 12 or 24 inputs.

In case of the CNN and the LSTM, we used the original signals $x_{t/b}$, $y_{t/b}$, and $p_{t/b}$ as input and allowed the networks to learn features from these signals by utilising one or multiple convolutional or LSTM layers (see Table S1). For the CNN, behind the convolutional layer(s), we used an average pooling layer to reduce the input dimension as commonly used for the convolutional network architectures. This was then followed by one or two FC layers. All details of the network architectures and their parameters can be found in Table S1. The schematic diagrams of an FFN architecture (F2), a CNN architecture (C3F2), and a LSTM architecture (L1F1.2) are presented in Figure 3(a), (b), and (c), respectively.

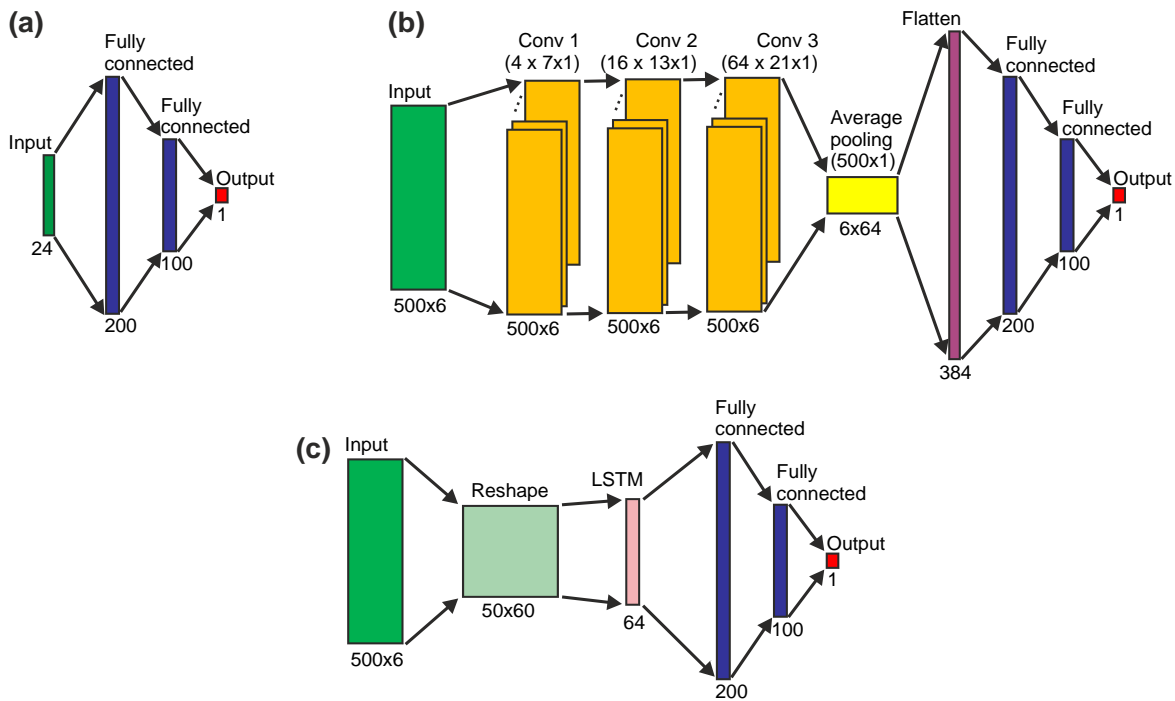

**Figure 3:** Schematic diagrams of the network architectures: **(a)** feed-forward network F2, **(b)** convolutional neural network C3F2, and **(c)** long short-term memory network L1F1.2. For more details, please see Table S1.

The SVMs were implemented using Python scikit-learn library[54]. We used either the radial basis function (RBF) kernels or the polynomial kernels of degrees 1-3. Regularisation parameter $C$ = [0.1, 1, 10, 100, 1000], and kernel coefficient gamma = [0.01, 0.1, 1, 10, 100] were tuned on the validation set (see section "Evaluation procedure and quantification measures" below).

All network architectures were implemented using TensorFlow[55] and Keras API[56]. To train the network architectures we used the Adam optimiser with a binary cross-entropy as a loss function, batch size 4, and default training parameters, i.e., learning rate = 0.001, $b_1$ = 0.9, $b_2$ = 0.999, and e = 1e-07. To avoid overfitting, we used a validation stop with validation split 1/6 and patience 10.

Data and source code is publicly available at https://doi.org/10.5281/zenodo.8104097.

**Evaluation Procedure and Quantification Measures**
To evaluate and compare the performances of the above presented classification models, we used a 5-fold cross-validation procedure. We divided the dataset into five subsets (each subset contained snippets from nine different infants). One subset was used as the test set for each fold, and the remaining four subsets were used to train the network architecture. The number of snippets in the training and test sets for each fold is given in Table 1. In each fold in the training set, we had on average 662 (SD = 4) snippets for the absence of fidgety movements (FM-) and 758 (SD = 3) for the presence of fidgety movements (FM+) class. In the test sets, we had on average 166 (SD = 4) snippets for the FM- and 190 (SD = 3) for the FM+ class.

The training set was split into training (5/6 of the training data) and validation (1/6 of the training data) subsets. In case of the SVM, we trained 25 models with different parameter combinations $C$ = [0.1, 1, 10, 100, 1000], and gamma = [0.01, 0.1, 1, 10, 100] on the training set. We then selected the model with the highest classification accuracy on the validation set, which was then evaluated on the test set. In case of the neural networks, for each fold we trained the network 20 times and then selected the model with the lowest loss score on the validation set, which was then evaluated on the test set.

For the evaluation of the classification performances, we used three common classification performance measures, i.e., sensitivity (true positive rate [TPR]), specificity (true negative rate [TNR]) and balanced accuracy (BA):

$$TPR = \frac{TP}{TP+FN}, TNR = \frac{TN}{TN+FP}, BA = \frac{TPR+TNR}{2},$$

where TP is the number of true positives, TN the number of true negatives, FP the number of false positives, and FN the number of false negatives.

**Table 1:** Data split of 5-fold cross-validation. The whole dataset contained 1776 snippets (828 FM- and 948 FM+) obtained from 45 infants. Each fold contained snippets from 36 and 9 infants for the training (~80% of snippets) and the test set (~20% of snippets), respectively. The training set was further split into the training (83.33% [30 infants]) and the validation (16.67% [6 infants]) subsets.

| Fold number | Training set (# of snippets) | | | Test set (# of snippets) | | |
|---|---|---|---|---|---|---|
| | FM- | FM+ | Total | FM- | FM+ | Total |
| 1 | 662 | 761 | 1423 | 166 | 187 | 353 |
| 2 | 662 | 754 | 1416 | 166 | 194 | 360 |
| 3 | 665 | 760 | 1425 | 163 | 188 | 351 |
| 4 | 666 | 757 | 1423 | 162 | 191 | 353 |
| 5 | 657 | 760 | 1417 | 171 | 188 | 359 |

**Results**

**Signal Examples**
Examples of the pressure mat sensor values and the extracted feature signals *x*, *y*, and *p* are shown in Figure 4, where the signals of an example of FM- and an example of FM+ are shown in the panels (a1, b1) and (a2, b2), respectively. In the panels (a1, a2), changes of pressure activity patterns caused by the infant's movement were presented. Extracted signals are shown in the panels (b1, b2). In case of an absence of fidgety movements (FM-, [b1]), local (short) signal patterns of lower frequency and larger amplitudes are observable. As a contrast to the FM+ case (b2), local (short) signal patterns of higher frequency and smaller amplitudes can be seen, which resembles the fidgety movement characteristics[27].

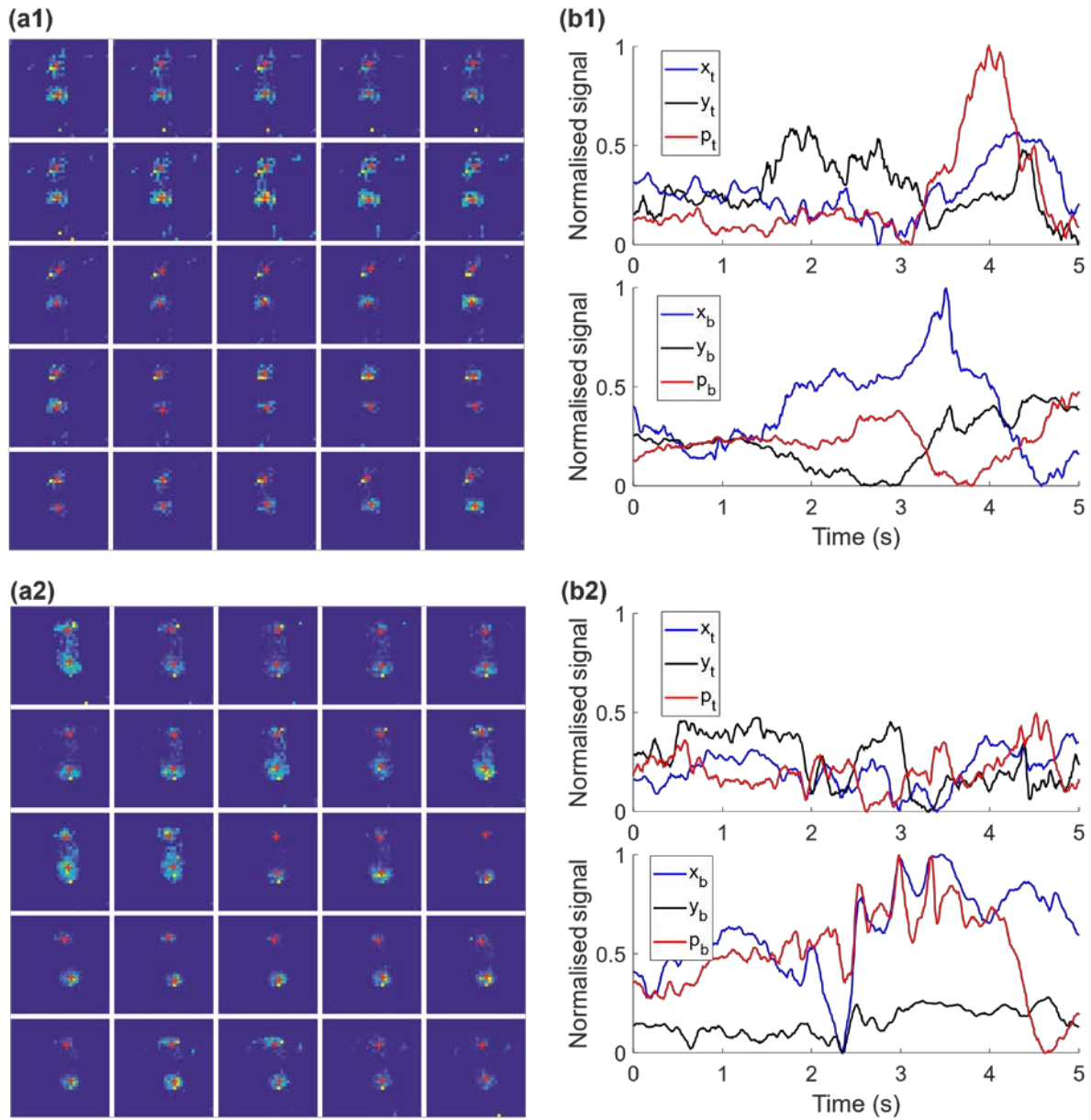

**Figure 4:** Examples of the pressure mat values and the extracted signals for one sample of the FM- (a1, b1) and one sample of the FM+ class (a2, b2). **(a)** Pressure mat values obtained at 0.2, 0.4, …, 5s. The blue and the red colours correspond to the low- and the high-pressure values, respectively. The red crosses correspond to the positions of the centre of mass for the top and the bottom areas of the pressure mat (see Figure 2). **(b)** Signals for position *x*, *y* of the centre of mass, and the average pressure *p*.

**Classification Results**

Results of the classification performances for the best models are summarised and compared in Figure 5. The performances of all models and all the performance measures are presented in Table S2.

**SVMs.** Applying SVMs with manually defined statistical features, the worst average classification performance was obtained when only the original signals *x*, *y*, and *p* (no derivatives, S1.RBF, S1.P1-3) were used, where the average balanced accuracy BA = 69.13-71.49%. Compared to the classification performance with additional features from the signal derivatives *x'*, *y'*, and *p'* (S2.RBF, S2.P1-3), the average BA = 73.87-**76.15%**. However, the improvement from 71.49% (S1.RBF) to 76.15% (S2.P1) was not statistically significant (*t-test, p = 0.0776*).

**FFN architectures.** Applying FFN architectures with manually defined statistical features, the worst average classification performance was obtained when only the original signals *x*, *y*, and *p* (no derivatives, network F1.1) were used, where the average balanced accuracy BA = 72.11%. Adding statistical features from the signal derivatives *x'*, *y'*, and *p'* (network F1.3) and increasing the number of neurons of a fully connected layer from 100 to 200 improved the classification performance (average BA = **75.57%**). However, this improvement was not statistically significant (*t-test, p = 0.2080*). Adding a second fully connected layer (network F2) did not improve the classification performance (average BA = 73.58%).

**CNN architectures.** A CNN network architecture with learned features and only one convolutional layer (C1F1.1, four filters of size 7x1) led to a better average classification performance (BA = **77.46%**), compared to the best average performance of the FFN architecture (BA = 75.57%). However, the difference was not statistically significant (*t-test, p=0.4305*).

Increasing the number of filters and the filter size (C1F1.2, 16 filters of size 13x1; C1F1.3, 64 filters of size 21x1) did not improve the classification performance, the average BA was 74.85% (C1F1.2) and 75.03% (C1F1.3), respectively. Increasing the number of neurons in a fully connected layer (C1F1.4) or adding a second fully connected layer (C1F2) also did not improve the classification performance, the average BA was 73.93% (C1F1.4) and 76.00% (C1F2), respectively.

Using architectures with two (C2F1) or three convolutional layers (C3F.1-2, C3F2) further improved the classification performance. The best classification performance was obtained by using a CNN architecture with three convolutional layers and two fully connected layers (C3F2), with an average BA = **81.43%**.

**LSTMs.** The classification performance using LSTMs was inferior than using the other classification models. The average BA of LSTMs ranged from 66.93% to 69.04%. The highest average classification accuracy was obtained using one LSTM layer and two FC layers, L1F2 (average BA = **69.04%**). Adding an additional LSTM layer and increasing the number of LSTM neurons did not improve the classification accuracy (average BA = 68.54%).

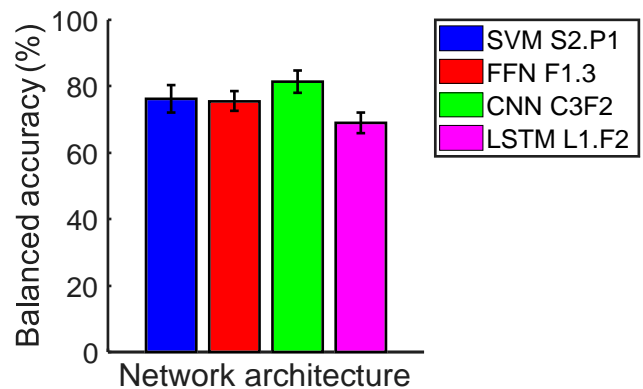

**Figure 5:** Comparison of the best performing network architectures (see Table S1) on their classification performances (FM+ vs. FM-). Coloured bars indicate for each model the average balanced accuracy (BA) obtained from the five test sets. Error bars denote the confidence intervals of the mean (CI 95%). The average BA between the CNN and the FFN, and between the CNN and the LSTM is significantly different (*t-test, p<0.05* in both cases). The average BA between the SVM and the CNN is not statistically significant (*t-test, p>0.05*).

**Model comparison.** Comparison of the best classification models of the four network architectures are shown in Figure 5. The CNN with learned features led to the highest average classification accuracy of 81.43% (CI = [78.00% 84.86%]). It outperformed the SVM (76.15%, CI = [72.00% 80.30%]) and the FFN (75.57%, CI = [72.65% 78.49%]) with manually defined features. The LSTM with learned features led to the worst classification performance (69.04%, CI = [65.94% 72.13%]).

## Discussion

In the current study, we carried out a proof-of-concept evaluation and explored the feasibility of using a pressure sensing device to track and classify age-specific infant general movement patterns. We adopted an existing pre-annotated dataset from a typically developing infant cohort[16] and examined whether a pressure data could be used to differentiate between typical movement patterns during the "fidgety age period" and the ones during the "pre-fidgety period". With the current pressure mat approach, and the convolutional neural network (CNN) architectures, the highest average classification accuracy achieved was 81%, with 86% sensitivity and 76% specificity for classifying *presence* vs. *absence* of the fidgety movements (FM+ vs. FM-).

We demonstrated that simple classification models such as support vector machines (SVMs) or feed-forward network architectures (FFN) with manually defined statistical features can reach a moderate classification accuracy (up to 76%). With the CNN architectures that allow for learning relevant features instead of predefining them, a higher classification accuracy (up to 81%) was achieved. The lowest accuracy was obtained using the long short-term memory (LSTM) network (69%).

### Classification performance (FM+ vs. FM-) of the pressure sensitive mat compared to the vision-based methods

While pressure sensitive devices have been used to evaluate infant sleep-wake behaviours, gross motor patterns, and postural control[40-50], they have rarely been used to classify infant general movements[51]. As such, at the moment, we can only compare the performance of the pressure mat to that of the vision-based sensors. We are aware that a direct comparison is not possible, since different data sets were used across different studies[9,11]. However, both the pressure mat technique and the vision-based methods are non-intrusive sensing approaches, which are conformable to the GMA guidelines[27]. In studies that attempted to distinguish FM+ versus FM-, the classification performance with the current pressure sensing mat seemed to be slightly lower than the performances based on RGB or RGB-D videos (see Table 2). The first possible reason for the lower performance might be that the mat used in this study was not specifically configured to capture infant motion. As technology advances continuously, more sensitive and suitable pressure sensing devices for infant motion tracking may improve the performance of the pressure-based classification. Second, the lower performance may be partially explained by the fact that the pressure sensing mat only measures motion of *some* body parts as compared to motion data obtained from full-body tracking. Infants during "active-wakefulness"[27] frequently lift their legs and arms above the lying surface[2], here, the pressure mat. The motions of the lifted extremities can therefore not be captured directly but are indirectly translated through the changing force distribution patterns of the body parts (e.g., head, shoulder, back, and hips) that are in contact with the mat. This may partly explain the lower classification performance of the pressure mat in the current study compared to our previous study using the same dataset but analysing full-body skeleton data[16]. Third, for this proof-of-concept study, we only randomly sampled a fraction of the available data of the entire data pool. The performance of the algorithm may be improved if a greater amount of data would be included in future studies. Adopting an existing expert-annotated dataset, the performance of the pressure mat was currently based on rather short, 5s long snippets. Using longer recordings for the machine learning procedures in the future might also improve the classification accuracy of the pressure sensing devices. Finally, one could also explore other CNN architectures such as temporal convolutional networks (TCNs) with residual connections which may improve the accuracy even further[57].

**The necessity of studying typical developmental patterns prior to atypical ones**

Importantly, data used in the current study originated from a healthy, typically developing cohort. The absence of fidgety movements (FM-) in this dataset reflects normal, age-typical motor patterns, i.e., the writhing movements[27], which are known to have significantly different quality and motion appearance from the pathological absence of fidgety movements[3]. The pathological absence-of-fidgety patterns of infants with neurological deficits, e.g., with monotonous, jerky, or cramped-synchronised movement characters[27], when compared to normal smooth and fluent fidgety movements of the same aged infants, could be easier to detect by the pressure mat and result in higher classification performance. Our current dataset however does not allow for testing this hypothesis. Rather, following a classic physiology-prior-pathology paradigm[58], with the data obtained from a healthy infant cohort, the present study was intended to examine the performance of the pressure sensing mat on classifying typical general movements, which are known to physiologically change their patterns during the first months of development, i.e., from the writhing to the fidgety pattern[27]. We would like to emphasise that the variabilities of the typical general movements are enormous, also within the same age period (e.g., the "fidgety age"), and shall not be underestimated[59]. Without a high-fidelity reflection and discernment of typical developmental patterns, attempts of classifying altered development may prove void. An AI-driven approach aiming at future clinical application therefore needs to investigate both typical and atypical patterns to warrant sensitivity as well as specificity.

**Pressure sensing approach – potential for generating public-accessible large datasets and safe data sharing aligning with privacy rules**

As has been discussed in the recent reviews on AI-based GMA approaches, the objectives and the applied datasets of the existing studies are dissimilar from each other, which makes direct comparison of diverse approaches difficult[9-11]. Different labs each work on a separate set of data, often limited in sample size. It was often not reported in the published works, how the respective dataset was annotated and validated[11], although valid annotations (e.g., FM+ vs. FM-) are the key for ML classification. Until now, no approved public-accessible large datasets are available in the field of researching general movements, although such dataset would be the basis for developing and comparing automated GMA solutions. This urgently calls for pooling and merging high-quality data across sites, which is challenging, partly due to the complex participants' confidentiality and privacy concerns[52,60]. With the pressure sensing data, participants' privacy protection can be easily achieved since no personal identifiable data is necessary for the analyses. As a comparison, with vision-based approaches, i.e., using cameras, which are predominantly used in the field[9-11], facial images, being one of the most sensitive personal identifiers, are commonly present in the datasets. Data de-identification and privacy protection can only be done through additional laborious technical manipulations[52].

**Advantages of the pressure sensing approach compared to other modalities for infant motion sensing**

Not to be forgotten, any clinical tool, if intended for broad application, has to be easy-to-use. Compared to other types of motion sensors (Table 3), data acquisition through pressure mats is non-intrusive, and requires minimum cost and setup efforts, which can be readily integrated into busy clinical routines almost anywhere. Importantly, the pressure sensing mat provides motion information in form of dynamic force changes in 2D-space. This is different from, and can provide a unique add-up to the information obtained through video data (usually body pose) or inertial motion sensors (acceleration and/or angular velocity)[28]. In case of cameras, additional algorithms need to be utilised to extract pose and/or motion information[14-16,61], whereas accelerometers inertial motion units (IMUs), or a pressure sensing mat provide motion information directly[20,23,24,46,51]. Although single RGB cameras are easy to install and operate (no synchronisation nor calibration required), they only provide 2D pose/motion information as compared to RGB-D cameras or IMU sensors. Both single RGB and RGB-D cameras frequently suffer from occlusions (either caused by the setup, e.g., in an incubator, or by the infant's own movements) and may lose track of some body parts from time to time. This limitation may be overcome by using multiple cameras[26,62]. However, such a setup becomes notably more complicated due to the necessity for synchronisation and calibration of the cameras, and also due to the amount of information generated, which needs to be processed to obtain 3D body pose and/or motion information[63,64]. Considering the strengths and limitations of different sensors[9-12,65], pressure sensing mat data can augment other sensor modalities, especially single RGB or RGB-D camera setups to improve infant motion analysis.

**Pressure sensing – a novel line of AI-driven GMA research and a vision of future development**
Our results demonstrate that the currently applied pressure mat, although not specifically designed for tracking infant movements, delivered promising classification results in distinguishing typical fidgety from pre-fidgety movements. Although the classification performance of the pressure mat, in common with that of the most existent automated GMA approaches, is still inferior than the performance of human GMA experts[10-12,65], the current study sets out an initial step for a novel line of non-intrusive AI-driven GMA research beyond using vision-based sensors (please also see below for comparison of different sensing modalities). It shall motivate further efforts to examine and improve the performance of the pressure sensors with extended datasets encompassing different patterns of general movements. Further studies shall evaluate if pressure sensors can reliably distinguish general movement patterns between: (a) normal pre-fidgety movements, i.e., age-typical writhing movement patterns before fidgety movements emerge; (b) abnormal pre-fidgety movements, e.g., poor-repertoire or cramp-synchronised patterns[27]; (c) normal movement patterns during the fidgety age period; and (d) abnormal movement patterns during the fidgety age period, e.g., absent or abnormal fidgety movements[27]. Our present study tested the classification between (a) and (c). This, from both the clinical and technological perspectives, is not comparable to the classification between (c) and (d), which was the focus of most studies listed in Table 2.

It has to be pointed out that the GMA is undoubtedly far beyond only identifying the presence or absence of the fidgety movements[3,11,66], although the fidgety movement pattern is of high diagnostic value[59] and has hence gained extensive attention, including in the field of developing automated GMA[10]. Naturally, motion information captured by the pressure sensing devices, like by many other sensors, can tell us much more than whether a specific motility (e.g., the fidgety movements) exists[29,32,35,43,47]. Future studies shall also investigate the potential of the pressure sensing mat for detecting other classes of motion features, within or beyond general movements, that are of clinical significance.

In short, pressure sensing solutions, by nature pseudonymised, provide a promising venue for realising easy and large-scale multi-centre data acquisition and sharing. This, if done, will enable synergy to develop, evaluate, and improve infant motion-tracking technologies that may ultimately scale up the implementation of AI-driven general movements assessment (GMA).

**Conclusion**

In this study, we demonstrated that pressure sensing methodology can generate adequate infant movement classification. With ongoing technological advances on infant-suitable devices, easy-to-apply, non-intrusive pressure sensing solutions have great potential to be applied in daily clinical practice and surveillance of infant neuromotor functions. Developing pressure sensing approaches will forcefully contribute to meeting the urgent need of acquiring and sharing large datasets across centres, and in return, accelerate further development and improvement of the technology. Considering pros and cons of different sensing techniques, we suggest that multimodal non-intrusive (i.e., pressure and video) data acquisition and analyses combining different venues of motion information may be a propitious direction for research and practice. This approach may optimise and streamline infant movement evaluations enabling efficient and broader clinical implementation of GMA and help to objectively identify infants at elevated likelihood for developing neuromotor disorders such as cerebral palsy.

**Table 2:** Comparison of classification performance of different methods for recognition of fidgety movements.

|  | Data | Classification/ Recognition | Classification performance measure ||||
|---|---|---|---|---|---|---|
|  |  |  | Sens. (%) | Spec. (%) | Balanced accuracy (%) | Accuracy (%) |
| **Kulvicius et al., current study** | Pressure sensing mat | FM+ vs. FM- | 86 | 76 | 81 |  |
| **Reich et al., 2021[16]** | RGB video | FM+ vs. FM- | 88 | 88 | 88 |  |
| **McCay et al., 2021[17]** | RGB-D video* | FM+ vs. FM- | 100 | 100 | 100 |  |
| **Tsuji et al., 2020[18]** | RGB video | FM recognition |  |  |  | 85 |
| **Machireddy et al., 2017[67]** | RGB video | FM recognition |  |  |  | 84 |
| **Adde et al., 2013[68]** | RGB video | FM recognition | 89 | 79 | 84 |  |
| **Adde et al., 2009[69]** | RGB video | FM+ vs. FM- | 90 | 80 | 85 |  |

*Synthetic MINI-RGBD dataset generated from RGB-D videos[70].

**Table 3:** Comparison of different sensors for acquisition of pose or motion information.

|  | Single RGB camera | Single RGB-D camera | Multiple RGB cameras | Multiple accelerometers/ IMUs | Pressure sensing mat |
|---|---|---|---|---|---|
| **Sensor type** | External (non-intrusive) | External (non-intrusive) | External (non-intrusive) | Wearable (on-body) | External (non-intrusive) |
| **Obtained pose/motion information** | 2D pose | 3D pose | 3D pose | 3D acceleration and 3D angular velocity | 2D position and pressure |
| **Extraction of pose/motion information** | Indirect | Indirect | Indirect | Direct | Direct |
| **Synchronisation required** | NO | NO | YES | YES | NO |
| **Calibration required** | NO | NO | YES | YES | NO |
| **Data privacy issue** | YES | YES | YES | NO | NO |
| **Applicability and handling in clinical settings** | Easy | Easy | Complicated | Complicated | Easy |

## Acknowledgements

The authors thank all families for their participation. Thanks to the team members of the Marschik Labs in Graz and Göttingen (iDN – interdisciplinary Developmental Neuroscience and SEE – Systemic Ethology and Developmental Science), who were involved in recruitment, data acquisition, curation, and pre-processing: Gunter Vogrinec, Magdalena Krieber-Tomantschger, Iris Tomantschger, Laura Langmann, Claudia Zitta, Dr. Robert Peharz, Dr. Florian Pokorny, Dr. Simon Reich. We are/were supported by BioTechMed Graz and the Deutsche Forschungsgemeinschaft (DFG – stand-alone grant 456967546, SFB1528 – project C03), the Laerdal Foundation, the Bill and Melinda Gates Foundation through a Grand Challenges Explorations Award (OPP 1128871), the Volkswagenfoundation (project IDENTIFIED), the LeibnizScience Campus, the BMBF CP-Diadem (Germany), and the Austrian Science Fund (KLI811) for data acquisition, preparation, and analyses. Special thanks also to our interdisciplinary international network of collaborators for discussing this study with us and for refining our ideas, and to the reviewers for their extensive feedback and advice.

# Supplemental material

**Table S1:** Details of the Support Vector Machine (SVM), Feed Forward Network (FFN), Convolutional Neural Network (CNN), and Long Short-Term Memory (LSTM) network architectures. After each fully connected (FC), convolutional layer (Conv) and LSTM layer, a batch normalisation layer, and a drop-out layer (10%) were used. ReLU activation functions were used in the FC, Conv and LSTM layers, whereas in the output layer a sigmoid activation function was used.

| SVM | | | | | | | |
|---|---|---|---|---|---|---|---|
| Name | Input dim. | Kernel | Output dimension | | | | |
| **S1.RBF** | 12 | Radial basis function | 1 | | | | |
| **S1.P1** | 12 | Linear polynomial | 1 | | | | |
| **S1.P2** | 12 | Quadratic polynomial | 1 | | | | |
| **S1.P3** | 12 | Cubic polynomial | 1 | | | | |
| **S2.RBF** | 24 | Radial basis function | 1 | | | | |
| **S2.P1** | 24 | Linear polynomial | 1 | | | | |
| **S2.P2** | 24 | Quadratic polynomial | 1 | | | | |
| **S2.P3** | 24 | Cubic polynomial | 1 | | | | |
| **FFN** | | | | | | | |
| Name | Input dim. | FC 1 neurons | FC 2 neurons | Output dimension | | | |
| **F1.1** | 12 | 100 | - | 1 | | | |
| **F1.2** | 24 | 100 | - | 1 | | | |
| **F1.3** | 24 | 200 | - | 1 | | | |
| **F2** | 24 | 200 | 100 | 1 | | | |
| **CNN** | | | | | | | |
| Name | Input dim. | Conv 1 filters / size | Conv 2 filters / size | Conv 3 filters / size | Average pooling | FC 1 neurons | FC 2 neurons | Output dim. |
| **C1F1.1** | 500x6 | 4 / 7x1 | - | - | 500x1 | 100 | - | 1 |
| **C1F1.2** | 500x6 | 16 / 13x1 | - | - | 500x1 | 100 | - | 1 |
| **C1F1.3** | 500x6 | 64 / 21x1 | - | - | 500x1 | 100 | - | 1 |
| **C1F1.4** | 500x6 | 64 / 21x1 | - | - | 500x1 | 200 | - | 1 |
| **C1F2** | 500x6 | 64 / 21x1 | - | - | 500x1 | 200 | 100 | 1 |
| **C2F1** | 500x6 | 4 / 7x1 | 16 / 13x1 | - | 500x1 | 100 | - | 1 |
| **C3F1.1** | 500x6 | 4 / 7x1 | 16 / 13x1 | 64 / 21x1 | 500x1 | 100 | - | 1 |
| **C3F1.2** | 500x6 | 4 / 7x1 | 16 / 13x1 | 64 / 21x1 | 500x1 | 200 | - | 1 |
| **C3F2** | 500x6 | 4 / 7x1 | 16 / 13x1 | 64 / 21x1 | 500x1 | 200 | 100 | 1 |
| **LSTM** | | | | | | | |
| Name | Input dim. | LSTM 1 neurons | LSTM 2 neurons | FC 1 neurons | FC 2 neurons | Output dim. | |
| **L1F1.1** | 25x120 | 64 | - | 100 | - | 1 | |
| **L1F1.2** | 50x60 | 64 | - | 100 | - | 1 | |
| **L1F1.3** | 100x30 | 64 | - | 100 | - | 1 | |
| **L1F1.4** | 50x60 | 64 | - | 200 | - | 1 | |
| **L1F2** | 50x60 | 64 | - | 200 | 100 | 1 | |
| **L2F2.1** | 50x60 | 64 | 32 | 200 | 100 | 1 | |
| **L2F2.2** | 50x60 | 128 | 64 | 200 | 100 | 1 | |

**Table S2:** Comparison of different classifiers (see Table S1) on classification of the presence vs. absence of fidgety movements. Average sensitivity, specificity and balanced accuracy is shown obtained on five test sets. Numbers in brackets correspond to confidence intervals of the mean (CI 95%).

| Classifier | Sensitivity (%) | Specificity (%) | Balanced Accuracy (%) |
|---|---|---|---|
| **SVM** | | | |
| **S1.RBF** | 73.15 [68.09 78.20] | 69.83 [62.39 77.27] | 71.49 [69.70 73.28] |
| **S1.P1** | 72.10 [66.46 77.73] | 69.95 [60.13 79.77] | 71.03 [68.77 73.28] |
| **S1.P2** | 72.30 [66.64 77.96] | 68.74 [58.94 78.53] | 70.52 [68.24 72.80] |
| **S1.P3** | 72.92 [68.40 77.43] | 65.35 [57.51 73.19] | 69.13 [66.63 71.63] |
| **S2.RBF** | 74.72 [70.68 78.77] | 73.01 [66.51 79.52] | 73.87 [70.92 76.82] |
| **S2.P1** | 78.60 [76.80 80.41] | 73.70 [65.65 81.75] | **76.15** [72.00 80.30] |
| **S2.P2** | 80.49 [77.56 83.43] | 71.54 [65.27 77.82] | 76.02 [72.71 79.32] |
| **S2.P3** | 79.04 [75.77 82.30] | 73.12 [67.39 78.85] | 76.08 [74.57 77.58] |
| **FFN** | | | |
| **F1.1** | 73.95 [70.60 77.29] | 70.27 [59.16 81.39] | 72.11 [68.11 76.11] |
| **F1.2** | 81.32 [78.46 84.17] | 68.29 [60.89 75.69] | 74.80 [71.16 78.45] |
| **F1.3** | 79.96 [77.11 82.82] | 71.18 [63.44 78.92] | **75.57** [72.65 78.49] |
| **F2** | 79.90 [74.02 85.78] | 67.25 [54.69 79.81] | 73.58 [69.52 77.63] |
| **CNN** | | | |
| **C1F1.1** | 85.35 [82.47 88.22] | 69.58 [62.03 77.13] | 77.46 [74.09 80.84] |
| **C1F1.2** | 81.66 [76.69 86.63] | 68.04 [58.18 77.90] | 74.85 [71.15 78.55] |
| **C1F1.3** | 80.57 [78.10 83.04] | 69.48 [64.00 74.96] | 75.03 [71.54 78.51] |
| **C1F1.4** | 78.03 [74.10 81.96] | 69.82 [62.59 77.05] | 73.93 [70.36 77.49] |
| **C1F2** | 79.26 [75.62 82.90] | 72.75 [65.08 80.41] | 76.00 [72.93 79.08] |
| **C2F1** | 84.42 [80.15 88.69] | 74.30 [62.43 86.16] | 79.36 [73.84 84.88] |
| **C3F1.1** | 84.48 [79.25 89.70] | 75.71 [68.62 82.80] | 80.09 [77.37 82.81] |
| **C3F1.2** | 86.06 [81.89 90.24] | 74.80 [67.72 81.89] | 80.43 [76.70 84.17] |
| **C3F2** | 86.48 [82.92 90.03] | 76.37 [70.71 82.04] | **81.43** [78.00 84.86] |
| **LSTM** | | | |
| **L1F1.1** | 69.54 [66.14 72.94] | 64.59 [56.34 72.83] | 67.06 [63.85 70.28] |
| **L1F1.2** | 77.89 [73.67 82.12] | 60.14 [50.85 69.42] | 69.01 [65.44 72.59] |
| **L1F1.3** | 76.10 [71.00 81.20] | 57.75 [50.24 65.25] | 66.93 [64.91 68.95] |
| **L1F1.4** | 75.11 [69.23 80.98] | 61.97 [51.86 72.08] | 68.54 [63.74 73.33] |
| **L1F2** | 78.13 [73.97 82.30] | 59.94 [53.60 66.29] | **69.04** [65.94 72.13] |
| **L2F2.1** | 82.21 [76.24 88.19] | 51.99 [34.45 69.54] | 67.10 [61.06 73.15] |
| **L2F2.2** | 75.11 [69.23 80.98] | 61.97 [51.86 72.08] | 68.54 [63.74 73.33] |